\documentclass[a4paper,fleqn]{cas-dc}
\usepackage[english]{babel}
\usepackage[numbers,sort&compress]{natbib}
\usepackage[dvipsnames]{xcolor}
\usepackage{amsmath, graphicx, hyperref}
\hypersetup{colorlinks=true, linkcolor=BurntOrange, citecolor=BurntOrange, filecolor=BurntOrange, urlcolor=BurntOrange}

\begin{document}

\title[mode=title]{New physics in toponium's shadow?}
\shorttitle{New physics in toponium's shadow?}

\let\printorcid\relax   
\author[first]{Thomas Flacke} 
\address[first]{Quantum Universe Center, KIAS, 85 Hoegi-ro, Dongdaemun-gu, Seoul 02455, Korea}
\author[second]{Benjamin Fuks} 
\address[second]{Laboratoire de Physique Théorique et Hautes Énergies (LPTHE), UMR 7589, Sorbonne Université et CNRS, 4 place Jussieu, 75252 Paris Cedex 05, France}
\author[third,fourth]{Dongchan Kim}

\address[third]{School of Physics, KIAS, 85 Hoegi-ro, Dongdaemun-gu, Seoul 02455, Korea}
\address[fourth]{Department of Physics, Korea University, Seoul 136-713, Korea}
\author[third]{Jinheung Kim} 
\author[first,third]{Seung J. Lee} 
\author[second]{Léandre Munoz-Aillaud} 
\shortauthors{T.~Flacke {\it et al.}}

\begin{abstract}
  ATLAS and CMS have recently reported enhancements in the top-antitop production rate near threshold, a region where non-perturbative QCD dynamics associated with toponium formation become relevant. We investigate how this behaviour is modified in the presence of a neutral pseudoscalar that couples to gluons and top quarks, using an effective description that consistently incorporates perturbative Standard Model and new physics contributions, their interference and non-perturbative threshold effects. We show that the combined effect of those ingredients markedly shapes the viable region of the pseudoscalar parameter space, particularly for narrow resonances with masses close to twice the top mass. While Standard Model threshold effects could explain a sizeable part of the measured enhancements, the current data remain compatible with additional contributions from pseudoscalar interactions.
\end{abstract}


\maketitle

\vspace*{-10.5cm}
  \noindent {\small\texttt{KIAS - Q25020}}
\vspace*{9.5cm}

\section{Introduction}

The production of top-antitop ($t\bar{t}$) pairs at the Large Hadron Collider (LHC) is a cornerstone of the Standard Model (SM) testing programme and a powerful probe of possible extensions of the SM. The kinematic threshold region, where the invariant mass of the top-antitop pair approaches twice the top quark mass $m_t$, offers a particularly sensitive environment. In this regime, observables depend not only on fundamental SM parameters but also on the intricate dynamics of QCD in its non-perturbative domain~\cite{Fadin:1987wz, Fadin:1988fn}. QCD indeed predicts the formation of quasi-bound $t\bar{t}$ states, collectively referred to as toponium. However, unlike the sharp resonances relevant to charmonium or bottomonium production, the top quark's large decay width implies that toponium states manifest as broad enhancements in the production cross section rather than narrow peaks~\cite{Fadin:1990wx, Strassler:1990nw, Jezabek:1992np, Sumino:1992ai, Hagiwara:2008df, Sumino:2010bv}.

For many years, the experimental observation of toponium effects at a hadron collider was considered extremely challenging. However, recent high-precision measurements by the ATLAS and CMS collaborations have revealed a robust and reproducible enhancement in the $t\bar{t}$ production cross section near threshold~\cite{ATLAS:2023fsd, CMS:2024pts, CMS:2024ybg, CMS:2024zkc, CMS:2025kzt, CMS:2025dzq, ATLAS:2025kvb}. Although the experiments do not claim evidence for toponium formation, the observed structures have reinvigorated interest in this long-anticipated QCD phenomenon. An early study pointed to toponium formation as a possible interpretation for the observed excesses~\cite{Fuks:2021xje}, and subsequently spurred numerous new analyses exploring this possibility~\cite{Maltoni:2024tul, Aguilar-Saavedra:2024mnm, Llanes-Estrada:2024phk, Fuks:2024yjj, Garzelli:2024uhe, Francener:2025tor, Nason:2025hix, Fuks:2025sxu, Bai:2025buy, Shao:2025dzw, Fuks:2025wtq}. In parallel, several works examined potential explanations in terms of physics beyond the SM (BSM)~\cite{Biekotter:2021qbc, Banik:2023vxa, Coloretti:2023yyq, Anuar:2024qsz, Maltoni:2024wyh, Lu:2024twj, Djouadi:2024lyv}, typically neglecting the role of toponium dynamics. However, as recently emphasised in~\cite{Fuks:2025toq}, a combined analysis systematically including both QCD bound-state effects and BSM contributions is still lacking, and in particular no study to date incorporates non-relativistic QCD (NRQCD) bound-state corrections directly into both the SM and BSM amplitudes despite the data consistently highlighting the possibility for toponium effects to coexist with BSM contributions.

The goal of this work is to fill this gap. Motivated by composite scenarios, we focus on BSM extensions featuring new scalar or pseudoscalar states with a mass near $2\, m_t$ and that dominantly couple to top quarks~\cite{Lee:1973iz, Kaplan:1983sm, Dugan:1984hq, Haber:1984rc, Georgi:1985nv, Chanowitz:1985ug, Branco:2011iw, Ferretti:2013kya, Cacciapaglia:2015eqa, Belyaev:2015hgo, Belyaev:2016ftv, BuarqueFranzosi:2018eaj, Cacciapaglia:2019bqz, Osipov:2019ccg, Cornell:2020usb}. While such a scalar or pseudoscalar mass is not generically predicted in weakly coupled extensions of the Standard Model, it naturally arises in certain strongly coupled or composite scenarios. In this context, a consistent interpretation of the LHC measurements in the top threshold regime should require incorporating the toponium effects not only in the (conventional) SM component of the related amplitude, but equally in the BSM one and its interference with the SM. Neglecting this interplay can induce distortions in the predicted threshold $t\bar t$ lineshape, leading to a mischaracterisation of the regions of the BSM parameter space compatible with the data and the phenomenological viability of BSM models embedding top-philic resonances.

To address this question, we perform fully realistic LHC simulations of $t\bar{t}$ production and decay that combine bound-state effects, SM and BSM contributions as well as their interference. Our analysis is based on a simple effective model in which a neutral pseudoscalar field couples to top quarks and gluons, that we implement in the \textsc{FeynRules} package~\cite{Christensen:2009jx, Alloul:2013bka, Darme:2023jdn} in order to be able to generate realistic Monte Carlo events with \textsc{MG5\_aMC}~\cite{Alwall:2014hca}. The contribution from toponium is then incorporated via the matrix-element re-weighting technique introduced in \cite{Fuks:2024yjj} which uses the Green's function of the non-relativistic QCD Hamiltonian. 

The rest of this paper is organised as follows. Section~\ref{sec:mnm} introduces the effective model and the method employed to incorporate toponium effects. Section~\ref{sec:BSMtopo} presents the combined impact of BSM and toponium contributions on $t\bar t$ production near threshold as well as its implications for the explored pseudoscalar parameter space. We conclude in Section~\ref{sec:con}.

\section{Model and Method}\label{sec:mnm}
Several ATLAS and CMS differential measurements of top-antitop production in the dileptonic channel~\cite{CMS:2024zkc, CMS:2025kzt, CMS:2025dzq, ATLAS:2025kvb} make use of observables which probe the spin structure of the associated scattering amplitudes through the angular correlations of the final-state charged leptons in their respective helicity frames~\cite{Bernreuther:2004jv, Bernreuther:2015yna}. Interestingly, both collaborations observe their largest deviations from the SM prediction in a phase space regime where contributions from CP-odd top-philic states are known to be enhanced. This motivates exploring SM extensions that include a pseudoscalar resonance coupling strongly to the top quark, and with a mass close to the $t\bar t$ threshold. Such pseudoscalar states arise naturally in many well-motivated BSM scenarios. A prominent example is provided by composite Higgs models with an underlying fermionic structure~\cite{Ferretti:2013kya, Cacciapaglia:2015eqa, Belyaev:2015hgo, Belyaev:2016ftv, BuarqueFranzosi:2018eaj, Cacciapaglia:2019bqz, Cornell:2020usb}, where pseudoscalars appear as pseudo-Nambu-Goldstone bosons with couplings determined by the representation of the strong-sector fermions and by the specific embedding of the top partners. Similar pseudoscalar degrees of freedom with sub-TeV masses and top-philic couplings also occur in supersymmetric models~\cite{Haber:1984rc}, in other composite scenarios of electroweak symmetry breaking~\cite{Kaplan:1983sm, Dugan:1984hq}, and in constructions with an extended scalar sectors such as multi-Higgs-doublet and custodial-triplet models~\cite{Lee:1973iz, Georgi:1985nv, Chanowitz:1985ug, Branco:2011iw}. 

\begin{figure}
    \centering
    \includegraphics[width=0.7\linewidth]{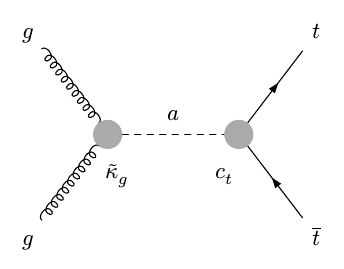}
    \caption{Feynman diagram representing the contribution of a pseudoscalar resonance $a$ to $t\bar t$ production in the gluon fusion channel. This amplitude interferes with the SM one and affects both the total rate and the spin-correlation structure.\label{fig:diagram}}
\end{figure}

We therefore consider a new-physics scenario in which we introduce a gauge-singlet pseudoscalar field $a$ of mass $M_a$ that couples only to gluons and top quarks, with interaction strengths governed by the two dimensionless parameters $\tilde\kappa_g$ and $c_t$. The corresponding effective Lagrangian reads
\begin{equation}\label{eq:lagrangian}\begin{split}
  \mathcal{L}_a =&\  \frac12 \partial_\mu a\, \partial^\mu a -\frac12 M_a^2\, a^2 +\frac{\alpha_s}{8\pi v}\tilde\kappa_g\, a\, G^a_{\mu\nu}\tilde{G}^{a\,\mu\nu} \\ &\  + 2 c_t\, a\, \bar t i\gamma_5 t\,,
\end{split}\end{equation}
where $v$ denotes the Standard Model Higgs vacuum expectation value and $\alpha_s$ the strong coupling constant. The effective coupling to gluons can receive two distinct contributions,
\begin{equation}\label{eq:kaapag_full}
  \tilde\kappa_g = \kappa_g + \kappa_t\,.
\end{equation}
The coupling $\kappa_g$ parametrises possible short-distance contributions from heavy new states, while $\kappa_t$ encodes the top-quark contribution generated by the axial anomaly. Even if $\kappa_g = 0$, a non-zero Yukawa coupling $c_t$ induces an $agg$ vertex at one loop, thereby generating the effective $a\,G\tilde{G}$ interaction relevant to $t\bar t$ production. The top-loop contribution $\kappa_t$ entering the $agg$ vertex is given by
\begin{equation}\label{eq:kappa_t}
  \kappa_t = \frac{v\, c_t}{2 m_t}\,  A_{1/2}^A\!\left(\frac{q^2}{4m_t^2}\right),
\end{equation}
where $A_{1/2}^A$ is the conventional pseudoscalar form factor defined, for example in~\cite{Djouadi:2005gj}, and $q$ stands for the pseudoscalar four-momentum. Strictly speaking, the $\kappa_t$ quantity from Eq.~\eqref{eq:kappa_t} corresponds to the one-loop contribution to the $agg$ vertex and not the full $aG\tilde{G}$ operator appearing in the Lagrangian~\eqref{eq:lagrangian}. However, at leading order in the process relevant for this work (see Figure~\ref{fig:diagram}), only the effective $agg$ vertex contributes, so it is sufficient to consider the top triangle loop contribution to $\tilde\kappa_g$.

The setup defined above is characterised by three free parameters, $M_a$, $c_t$ and $\kappa_g$. In order to simplify the discussion and to enable a transparent exploration of the parameter space, we treat $M_a$ and $c_t$ as the independent inputs, while fixing $\kappa_g$ through the total width of the pseudoscalar. Concretely, we consider two benchmark scenarios in which the relative pseudoscalar width is set to $\Gamma_a/M_a = 1\%$ or $5\%$, and we determine $\kappa_g$ such that the pseudoscalar acquires the chosen width-to-mass ratio after assuming that the dominant decay modes are $a \to gg$ and $a \to t\bar t$ when kinematically open. In this way, the parameter $\kappa_g$ is fully determined up to a sign flip for a given set of parameters $(c_t,M_a, \Gamma_a)$, and we choose throughout this work the positive solution. For practical calculations, we employ a public model implementation in {\sc FeynRules}~\cite{Christensen:2009jx, Alloul:2013bka, Darme:2023jdn}\footnote{See the `\textsc{eVLQ S012}' model file available from \url{http://feynrules.irmp.ucl.ac.be/wiki/NLOModels}.}, which we modify to turn off all irrelevant couplings and include the changes induced by Eqs.~\eqref{eq:kaapag_full} and \eqref{eq:kappa_t} and then export it to \textsc{MG5\_aMC}~\cite{Alwall:2014hca} for cross section calculations.

The toponium contribution is incorporated following the matrix-element re-weighting strategy introduced in~\cite{Fuks:2021xje, Fuks:2024yjj, Fuks:2025wtq}. In this approach, we project the partonic amplitude onto a colour-singlet configuration and multiply the resulting fixed-order matrix element by the ratio of the NRQCD S-wave Green's function $\widetilde G(E,p^*)$ to the free one $\widetilde G_0(E,p^*)$,
\begin{equation}\label{eq:reweight}
|\mathcal{M}|^2 \;\to\; \left| \mathcal{M}\,  \frac{\widetilde G(E,p^*)}{\widetilde G_0(E,p^*)}\right|^2\,,
\end{equation}
thus embedding Coulombic resummation effects into the event generation process.  Here, $E$ denotes the toponium binding energy and $p^*$ is the common magnitude of the top and antitop three-momenta in the $t\bar t$ rest frame. We employ the Green's function in momentum space $\widetilde G(E,p)$ obtained by solving the Lippmann-Schwinger equation with the Fourier transform of a leading-order QCD potential~\cite{Jezabek:1992np, Hagiwara:2016rdv, Hagiwara:2017ban}, and that is publicly provided as a lookup table.\footnote{See \url{http://github.com/BFuks/toponium.git}.} The re-weighting procedure is finally applied only in the kinematic window where NRQCD is expected to hold, which corresponds approximately to $t\bar t$ invariant masses $m_{t\bar t}\lesssim 350\,$GeV and to $p^*\lesssim 50\,$GeV.

Under this description of the toponium effects, we model the total $t\bar{t}$ production rate as the sum of two contributions,
\begin{equation}\label{eq:sigmasum}
  \sigma = \sigma_{\text{pQCD}} + \sigma_{\text{NRQCD}}\,.
\end{equation}
Here $\sigma_{\text{pQCD}}$ denotes the purely perturbative prediction that we evaluate at leading order (LO) in QCD, while $\sigma_{\text{NRQCD}}$ represents the additional threshold contributions obtained from matrix-element re-weighting based on the Green's function of the NRQCD Hamiltonian. Both $\sigma_{\text{pQCD}}$ and $\sigma_{\text{NRQCD}}$ are computed using the {\sc MG5\_aMC} generator (version~3.6.3), with parton luminosities evaluated from the LO NNPDF23 parton distribution functions~\cite{Ball:2013hta}. 

The above formalism can be employed both in the SM and in the presence of the pseudoscalar state described by the Lagrangian~\eqref{eq:lagrangian}. In this last case, the partonic matrix element is given by the coherent sum
\begin{equation}\label{eq:matrixelementfull}
    \mathcal{M}_{\mathrm{BSM}} = \mathcal{M}_{\mathrm{QCD}} + \mathcal{M}_{a}\,,
\end{equation}
where $\mathcal{M}_{\mathrm{QCD}}$ denotes the SM contribution and $\mathcal{M}_{a}$ the amplitude induced by the diagram of Figure~\ref{fig:diagram}. Since the NRQCD re-weighting  of Eq.~\eqref{eq:reweight} is applied directly to the squared matrix element, we thus respectively consider two different replacements for the SM and BSM cases,
\begin{equation}\label{eq:reweightfull}\begin{split}
    |\mathcal{M}_{\mathrm{QCD}}|^2 \;\to\; &\ \left| \mathcal{M}_{\mathrm{QCD}}\,\frac{G(E,p^*)}{G_0(E,p^*)} \right|^2\,,\\[.2cm]
    |\mathcal{M}_{\mathrm{BSM}}|^2 \;\to\; &\ \left| \mathcal{M}_{\mathrm{BSM}}\,\frac{G(E,p^*)}{G_0(E,p^*)} \right|^2\,,
\end{split}\end{equation}
which explicitly shows that any modification of the short-distance dynamics affects both the perturbative and the toponium-induced components of the $t\bar{t}$ cross section. In principle, the exchange of the pseudoscalar $a$ also generates a short-range non-relativistic potential between the top quarks. Its range is however exponentially suppressed by $\exp(-M_a r)$, so the resulting potential is subdominant compared to the QCD Coulomb interaction \cite{Daido:2017hsl, Fadeev:2022wzg} and we therefore neglect this effect. To streamline the discussion of the various ingredients entering our calculation, we introduce the auxiliary cross sections of Table~\ref{tab:xsec}. The quantities $\sigma_{\mathrm{pQCD}}$ and $\sigma_{\mathrm{pBSM}}$ are purely perturbative LO cross sections, evaluated respectively with $|\mathcal{M}_{\mathrm{QCD}}|^2$ and $|\mathcal{M}_{\mathrm{BSM}}|^2$. Their NRQCD-enhanced counterparts, $\sigma_{\mathrm{QCD}}$ and $\sigma_{\mathrm{BSM}}$, additionally include the contribution $\sigma_{\mathrm{NRQCD}}$ obtained by integrating the re-weighted squared matrix elements in Eq.~\eqref{eq:reweightfull} over the non-relativistic region. In other words,
\begin{equation}\begin{split}
  \sigma_{\mathrm{QCD}} = &\ \sigma_{\mathrm{pQCD}} + \sigma_{\mathrm{NRQCD}}[\mathcal{M}_{\mathrm{QCD}}]\,,\\
  \sigma_{\mathrm{BSM}} =&\ \sigma_{\mathrm{pBSM}} + \sigma_{\mathrm{NRQCD}}[\mathcal{M}_{\mathrm{BSM}}]\,,
\end{split}\end{equation}
\begin{table}\renewcommand{\arraystretch}{1.5}
  \centering \begin{tabular}{c|l}
  Cross section & Short-distance matrix element used \\
  \hline
    $\sigma_{\mathrm{pQCD}}$ & $|\mathcal{M}_{\mathrm{QCD}}|^2$ \\
    $\sigma_{\mathrm{pBSM}}$ & $|\mathcal{M}_{\mathrm{BSM}}|^2$ \\
    $\sigma_{\mathrm{QCD}}$  & $|\mathcal{M}_{\mathrm{QCD}}|^2$ and its re-weighted counterpart \\
    $\sigma_{\mathrm{BSM}}$  & $|\mathcal{M}_{\mathrm{BSM}}|^2$ and its re-weighted counterpart \\
  \end{tabular}
  \caption{Summary of the different cross sections used in our study. The NRQCD contributions are obtained from the re-weighted matrix elements according to Eq.~\eqref{eq:reweightfull}, together with a phase space integration limited to the non-relativistic regime.\label{tab:xsec}}
\end{table}

For the $\sigma_{\text{NRQCD}}$ computations specific to this study, we modify the re-weighting procedure of~\cite{Fuks:2021xje, Fuks:2024yjj, Fuks:2025wtq} in a twofold way. First, since we focus only on total production rates (see Section~\ref{sec:BSMtopo}), we consider the $2\to 4$ process $gg\to b W^+ \bar b W^-$ without simulating the decays of the $W$ bosons. Second, we update differently the associated colour matrix to project the intermediate $t\bar t$ system onto a colour-singlet toponium state. In the presence of BSM contributions, this colour matrix is a $3\times3$ matrix, with the first $2\times2$ block corresponding to the SM components, the $(3,3)$ element to the BSM contribution and the off-diagonal elements to the interference between the SM and BSM diagrams. Subsequently, the singlet projection has only to be applied to the SM block as the pseudoscalar is a colour-singlet already, which leads to the replacement\footnote{Since MadGraph v3.6.2, the colour matrix is stored as an upper-triangular matrix with an additional symmetry factor of 2 to be added for the off-diagonal elements.}
\begin{equation}
\frac{1}{3}
\begin{pmatrix}
16 & -2 & 6 \\
-2 & 16 & 6 \\
6 & 6 & 18
\end{pmatrix}
\quad\longrightarrow\quad
\frac{1}{3}
\begin{pmatrix}
2 & 2 & 6 \\
2 & 2 & 6 \\
6 & 6 & 18
\end{pmatrix}\,.
\end{equation}
This modification ensures that only colour-singlet toponium configurations contribute to the NRQCD-re-weighted cross section, while the interference between the SM and BSM amplitudes is correctly preserved.


\begin{figure*}
  \includegraphics[width=\linewidth,trim={2.2cm 0 2.5cm 0.5cm},clip]{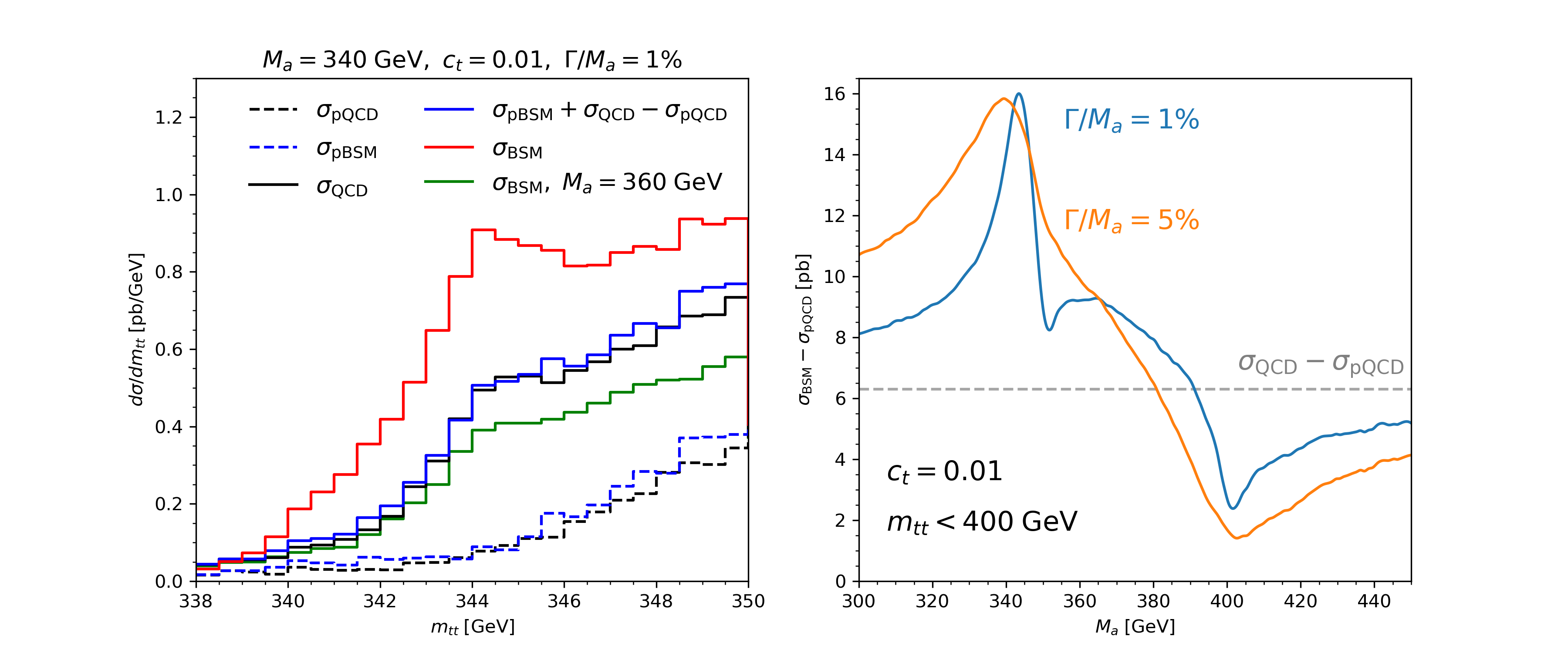}
  \caption{Invariant mass $m_{t\bar t}$ distribution (\textit{left}) and total cross-section shifts (\textit{right}) induced by the presence of the pseudoscalar state in the theory. \textit{Left} -- Predictions for a scenario with $c_t = 0.01$ and $M_a = 340$~GeV. We show perturbative SM (dashed black) and perturbative BSM (dashed blue) results, together with predictions including toponium effects through NRQCD matrix-element re-weighting (solid black and solid red, respectively). These are compared to the naive combination of perturbative BSM rates with SM toponium effects (solid blue), and to results for $M_a = 360$~GeV (solid green). \textit{Right} -- Dependence of $\sigma_{\mathrm{BSM}}-\sigma_{\mathrm{pQCD}}$ on $M_a$ for $c_t = 0.01$, $m_{t\bar t}<400$~GeV and assuming width-to-mass ratios of 1\% (blue) and 5\% (orange). The predictions are compared with the toponium expectation from NRQCD (dashed grey).\label{fig:mtt}}
\end{figure*}

\section{When New Physics Meets Toponium at the Top-Antitop Threshold}\label{sec:BSMtopo}
Having established how to obtain reliable predictions for top-antitop production near threshold in the presence of new physics using \textsc{MG5aMC}, we now turn to the associated phenomenological consequences. We begin by examining how a pseudoscalar state modifies the $m_{t\bar t}$ spectrum in the threshold region, where non-perturbative Coulomb effects are largest, and by demonstrating why a consistent treatment rather than a naive addition of perturbative ingredients is essential, as emphasised in~\cite{Fuks:2025toq}.

In Figure~\ref{fig:mtt}, we present the differential $t\bar t$ production cross section as a function of $m_{t\bar{t}}$ for a benchmark scenario defined by $M_a = 340$~GeV, $c_t = 0.01$ and $\Gamma/M_a = 1\%$. The perturbative QCD prediction $\sigma_{\mathrm{pQCD}}$ (dashed black) in which only SM diagrams are included is compared to the full QCD result $\sigma_{\mathrm{QCD}}$ including toponium effects (solid black). The $\sigma_{\mathrm{QCD}}$ cross section exhibits an enhancement below 350~GeV, which corresponds to the manifestation of the Coulombic QCD interactions between the slowly moving non-relativistic top and antitop quarks that are captured by the NRQCD Green's function. For the chosen small BSM coupling $c_t$, the perturbative BSM predictions $\sigma_{\mathrm{pBSM}}$ (dashed blue) is numerically very close to the $\sigma_{\mathrm{pQCD}}$ one. Consequently, simply adding the SM toponium correction to the perturbative BSM result, which corresponds to $\sigma_{\mathrm{pBSM}} + \sigma_{\mathrm{NRQCD}}[\mathcal{M}_{\mathrm{QCD}}]$, yields a curve (solid blue) almost indistinguishable from the $\sigma_{\mathrm{QCD}}$ one. By contrast, the full result $\sigma_{\mathrm{BSM}}$ including NRQCD matrix re-weighting applied to the complete perturbative BSM amplitude using $\sigma_{\mathrm{NRQCD}}[\mathcal{M}_{\mathrm{BSM}}]$ for the toponium component (solid red) differs noticeably. This behaviour reflects the colour structure of the underlying amplitudes: while the SM $gg\to t\bar{t}$ process receives both colour-singlet and colour-octet contributions, the pseudoscalar diagram produces the $t\bar{t}$ pair exclusively in a singlet state. As a result, the non-perturbative enhancement from the Green's function is significantly more pronounced than one would infer from a naive perturbative comparison.

In addition, it is known that the interference between the SM and pseudoscalar contributions near $m_{t\bar{t}} \simeq M_a$ strongly impact the differential cross section, with constructive or destructive patterns depending on the precise mass and width of the pseudoscalar state. While these interference effects in the perturbative regime have been emphasised for a long time~\cite{Gaemers:1984sj, Dicus:1987fk, Dicus:1994bm, Bernreuther:1997gs}, our results show that their interplay with the threshold-enhanced NRQCD contribution leads to a qualitatively new conclusion, highlighting the necessity of combining perturbative and non-perturbative physics consistently. To illustrate such an impact, the green curve in Figure~\ref{fig:mtt} shows the corresponding prediction for a heavier pseudoscalar with mass $M_a=360$~GeV. In this case, the cross section is reduced relative to the full SM prediction due to destructive interference between the SM and BSM amplitudes. Again, the deviation is larger than one would expect from perturbative information alone, which illustrates the strong sensitivity of the threshold region to the value of $M_a$: even a small shift in the pseudoscalar mass significantly changes the overlap with the toponium-enhanced region.

To further illustrate the dependence of the full cross section on the pseudoscalar mass, the right panel of Figure~\ref{fig:mtt} shows the difference $\sigma_{\mathrm{BSM}}-\sigma_{\mathrm{pQCD}}$ as a function of $M_a$ after applying a parton-level cut of $m_{t\bar t}<400$~GeV. The blue and orange curves correspond to scenarios featuring a relative pseudoscalar width of $1\%$ and $5\%$, respectively, while the dashed grey line corresponds to 6.43~pb and indicates the SM toponium contribution as predicted by NRQCD~\cite{Sumino:2010bv, Fuks:2021xje}. For the narrow width case ($\Gamma/M_a=1\%$), the presence of the pseudoscalar enhances the production rate over the SM prediction. As $M_a$ increases, the enhancement grows and reaches a maximum around $M_a\simeq340$~GeV, close to the toponium region, while around $M_a\simeq 2 m_t$, the SM-BSM interference switches sign (cf.\ Eq.~\eqref{eq:kappa_t}), producing the local minimum observed in the blue curve. At higher masses, $\sigma_{\mathrm{BSM}}$ decreases, and beyond $M_a\approx350$~GeV a narrow pseudoscalar no longer contributes significantly to toponium formation but continues to affect the perturbative $gg\to t\bar t$ rate. Finally, for $M_a\gtrsim400$~GeV, the pseudoscalar resonance lies outside the $m_{t\bar t}<400$~GeV bin and its effect in the threshold regime is effectively suppressed. The broader scenario ($\Gamma/M_a=5\%$) exhibits a similar peak-dip pattern, but the structure is smeared due to the larger pseudoscalar width. This illustrates how the resonance width controls the sharpness of the threshold enhancement: larger widths dilute the contribution over a wider $m_{t\bar{t}}$ range, reducing the maximum deviation and smoothing the peak-dip structure.

We now assess the phenomenological impact of combining perturbative SM and BSM amplitudes with non-relativistic corrections applied to the full $t\bar t$ production matrix element to constrain the pseudoscalar parameter space. The key question addressed is whether the ATLAS and CMS observed excesses of events near the $t\bar t$ threshold can be accommodated within the SM only, or whether it could point toward additional contributions from a light pseudoscalar.

Both collaborations have recently measured the $t\bar{t}$ production rate in the dilepton channel using dedicated selections targeting the threshold region. In all cases, the extracted cross sections are found to lie above the perturbative SM expectation~\cite{CMS:2024ybg, CMS:2025kzt, CMS:2025dzq, ATLAS:2025kvb}. The different analyses, however, differ in their $m_{t\bar t}$ binning choices and in the treatment of the non-relativistic toponium component. To remain conservative and aligned with our own simulation pipeline, we adopt the ATLAS determination as a benchmark~\cite{ATLAS:2025kvb}, the observed excess consistent with a non-relativistic toponium contribution being
\begin{equation}
    \sigma_\mathrm{obs}^{\mathrm{ATLAS}}(t\bar{t}_{\mathrm{NRQCD}}) = 9.0 \pm 1.3~\mathrm{pb}\,,
\end{equation}
with the excess localised in the region $m_{t\bar{t}} < 400~\mathrm{GeV}$. Since this quantity effectively corresponds to the difference between the data and the SM perturbative QCD prediction $\sigma_\mathrm{pQCD}$, an allowed point in the pseudoscalar parameter space must satisfy, at the $k\sigma$ level,
\begin{equation}\begin{split}
    \Delta\sigma = &\ \sigma_\mathrm{BSM} - \sigma_\mathrm{pQCD} \in \big(9.0 \pm 1.3k\big)\,\mathrm{pb}\\ 
    &\hspace{.4cm} \qquad\text{for}\ \ m_{t\bar{t}} < 400~\mathrm{GeV}\,.
\end{split}\end{equation}
This approach has the advantage of capturing the observed enhancement of threshold production while remaining agnostic to the phase-space-dependent implementation details of the ATLAS analysis. Here, recall that $\sigma_\mathrm{BSM}$ includes both the SM and BSM perturbative contributions as well as the non-perturbative corrections described in Section~\ref{sec:mnm}. By contrast, the naive approach fully ignoring bound-state effects would constrain the BSM parameter space from the quantity
\begin{equation}
    \Delta\hat\sigma = \sigma_\mathrm{pBSM} - \sigma_\mathrm{pQCD}\,,
\end{equation}
whereas ignoring the BSM contributions to the non-perturbative enhancement from the Green's function would rely instead on
\begin{equation}
    \Delta\tilde\sigma = (\sigma_\mathrm{pBSM} -\sigma_\mathrm{pQCD}) + (\sigma_\mathrm{QCD}-\sigma_\mathrm{pQCD})\,.
\end{equation}

\begin{figure*}
  \centering
  \includegraphics[width=1\linewidth]{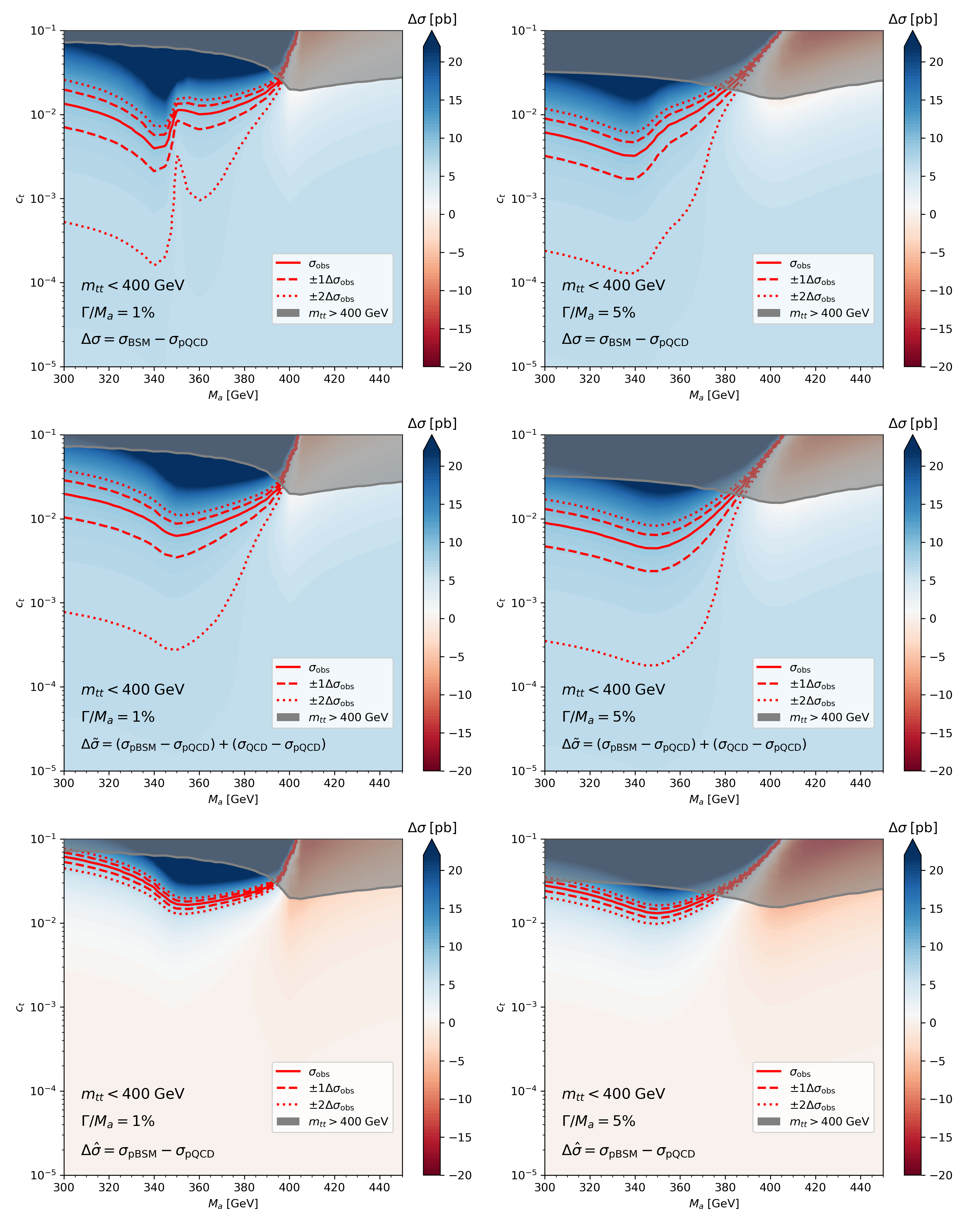}
  \caption{Exclusion contours in the $(M_a, c_t)$ plane for $\Gamma/M_a = 1\%$ (left) and $5\%$ (right). The three rows correspond to different theoretical assumptions: the full perturbative SM and BSM contributions supplemented by the non-perturbative corrections (top row), predictions in which the non-perturbative corrections are applied only to the SM part (middle row), and predictions with only perturbative contributions (bottom row).\label{fig:exclusion}}
\end{figure*}

Figure~\ref{fig:exclusion} summarises the resulting constraints in the $(M_a, c_t)$ plane for a pseudoscalar of relative width $1\%$ (left) and $5\%$ (right). In all cases, $\kappa_g$ is fixed by the chosen width, and the solid, dashed and dotted red lines denote the central, $1\sigma$ and $2\sigma$ contours of the allowed region of the pseudoscalar parameter space. To ensure consistency with high-mass data where no excess is observed, we additionally impose
\begin{equation}
    \sigma_\mathrm{BSM} < 1.05\,\sigma_\mathrm{pQCD} \quad\text{for}\ \ m_{t\bar{t}} > 400~\mathrm{GeV}\,,
\end{equation}
with the corresponding excluded region being shaded in grey in each panel of Figure~\ref{fig:exclusion}. 

When ignoring any threshold effects, the entire excess must be attributed to perturbative BSM production, so $\Delta\hat\sigma \in (9.0 \pm 1.3k)$~pb. The resulting viable region of the parameter space is shown in the bottom row of the figure. We find that reaching agreement between the data and the theoretical predictions necessitates a comparatively large value of $c_t$ for $M_a \lesssim 2m_t$, even if the SM-BSM interference is constructive in this parameter space region. For $M_a \gtrsim 2m_t$ the interference becomes destructive, yet a smaller coupling is sufficient to reproduce the excess. This behaviour stems from the fact that, above threshold, the pseudoscalar can be produced on shell, which significantly enhances its contribution despite the destructive interference. This dependence weakens for larger widths where the interference pattern is effectively smeared out. In a second scenario (middle row) where the non-perturbative effects are applied only to the SM part (\textit{i.e.}\ we impose $\Delta\tilde\sigma\in (9.0 \pm 1.3k)$~pb), the viable region of the parameter space indicates that the SM non-perturbative contribution alone almost saturates the observed enhancement but lies slightly below it. Consequently, moderate values of $c_t$ are preferred, and the dependence on $M_a$ follows the expected interference pattern. Importantly, this implicitly assumes that the pseudoscalar does \emph{not} modify the threshold dynamics beyond its perturbative amplitude. In contrast, when the full prediction with BSM-modified toponium contributions is used (top row), the allowed parameter space region exhibits qualitatively new features whenever $M_a \approx 2m_t$. In this region, the presence of the pseudoscalar both alters the interference pattern and toponium formation, leading to a pronounced restructuring of the allowed band. These distortion are most visible for narrow widths, while for larger widths these effects are smoothed out but not entirely eliminated. 

Overall, the combined effect of interference and threshold dynamics reduces the size of the viable coupling values compared to the perturbative-only scenario, and significantly reshapes the exclusion contour near $M_a\simeq 2m_t$. Additionally, these features highlight the necessity of a consistent treatment of BSM-modified toponium production when deriving constraints from threshold observables.

\section{Conclusion}\label{sec:con}

ATLAS and CMS have recently reported excesses in the $t\bar t$ production rate near threshold. While a moderate enhancement is expected from SM non-perturbative dynamics associated with toponium formation, the impact of the measured excesses seem to be larger than what is obtained from the SM alone and motivates a closer examination of the threshold region. In this work we have performed a consistent study of $pp \to t\bar t$ production including perturbative SM and pseudoscalar BSM amplitudes, their interference and toponium non-perturbative effects modelled through the re-weighting of the corresponding matrix elements by the NRQCD Green's function. Our setup thus provides a unified prediction for the threshold cross section in which BSM-induced modifications and bound-state effects are treated on the same footing.

Our results show that SM threshold effects account for a sizeable part of the observed enhancement, but do not preclude an additional contribution from pseudoscalar interactions. The viable parameter space is found to be strongly shaped by the interplay between the SM-BSM interference and the threshold dynamics, especially for narrow resonances with masses close to $2m_t$. In this case, the BSM state indeed modifies both the short-distance kernel and the formation of the toponium system. 

Although our analysis has focused on the inclusive threshold rate, the framework can be extended to differential observables and to broader classes of models, including two-Higgs-doublet, axion-like or custodial multiplet scenarios. Given the sensitivity of the threshold observables to both perturbative and non-perturbative effects, improved measurements and a systematic treatment of BSM-modified threshold dynamics will be essential to determine whether the observed enhancements are fully attributable to SM toponium formation, or whether they hint at new interactions of the top quark.

\section*{Acknowledgments}
This work has been supported  by the Center for Advanced Computation at Korea Institute for Advanced Study and the Korea-France Science and Technology Amicable Relationships (STAR) programme (MEST No.~RS-2023-00259500 for the Korean side and project No.~50136PE for the French side). TF is supported by a KIAS Individual Grant (QP083701) via the Quantum Universe Center at the Korea Institute for Advanced Study. BF is supported in part by Grant ANR-21-CE31-0013 (Project DMwithLLPatLHC) from the French \emph{Agence Nationale de la Recherche} (ANR). JK has been supported by a KIAS Individual Grant (PG099201) at Korea Institute for Advanced Study. SJL is supported in part by the Samsung Science Technology Foundation under Project No. SSTF-BA2201-06.

\bibliographystyle{JHEP}
\bibliography{literature}

\end{document}